# Chemical Self Assembly of Graphene Sheets

Hailiang Wang, Xinran Wang, Xiaolin Li and Hongjie Dai[*]

*Department of Chemistry and Laboratory for Advanced Materials, Stanford University, Stanford, CA 94305, USA*

*\* Correspondence to hdai@stanford.edu*

**ABSTRACT**

Chemically derived graphene sheets (GS) were found to self-assemble onto patterned gold structures via electrostatic interactions between noncovalent functional groups on GS and gold. This afforded regular arrays of single graphene sheets on large substrates, characterized by scanning electron and Auger microscopy (SEM) imaging and Raman spectroscopy. Self assembly was used for the first time to produce on-substrate and fully-suspended graphene electrical devices. Molecular coatings on the GS were removed by high current "electrical annealing", which recovered the high electrical conductance and Dirac point of the GS. Molecular sensors for highly sensitive gas detections are demonstrated with self-assembled GS devices.

Producing arrays of novel materials with ordered structures as device building blocks represents an important approach to nanoelectronics [1-4]. Graphene has recently shown interesting properties and application potentials including quantum Hall effect [5], field effect transistors [6, 7], sensors [8] and transparent electrodes [9]. The making of graphene has been achieved by mechanical peeling-off [10] and epitaxial growth [11]. Recently chemical methods [12-15] have attracted much attention due to high yield and scalability. Thus far, graphene sheets (GS) made chemically were typically deposited randomly on substrates. It is highly desirable to assemble GS at specific locations and into desired patterns, for rational design of functional GS devices at a large scale. Only one report exists thus far for assembly of highly insulating graphite oxide (GO) [16] on metal surfaces.

Here, we report self-assembly of highly-conducting graphene sheets driven by electrostatic interactions between noncovalent functional groups on GS and gold surfaces. We found that GS functionalized by tetrabutylammonium (TBA) and phospholipid-polyenthyleneglycol-amine (PL-PEG-NH$_2$) spontaneously adsorbed onto patterned gold surfaces. Auger elemental mapping/imaging, scanning electron microscopy (SEM) and Raman mapping revealed high selectivity of GS deposition on gold over silicon oxide substrate. Arrays of single GS on substrates were obtained this way. By self-assembly, graphene sheets were directed to bridge arrays of gold electrodes on substrate, affording large numbers of on-substrate GS devices and fully-suspended devices. An "electrical annealing" method was used to remove the coatings on GS, increase the electrical conductance of GS device and make the GS doping-free for applications such as chemical sensors.

Large quantities of high quality GS (sheet size ~200nm-1μm) were synthesized by an exfoliation-reintercalation-expansion method [14]. In a typical synthesis (see Experimental), ground natural graphite was intercalated by oleum in the presence of sodium nitrate, and then the product was inserted by TBA hydroxide aqueous solution and suspended by PL-PEG-amine in N, N-dimethylformamide (DMF) [17]. The final GS solvent was $H_2O$/DMF with a volume ratio of ~3/100. Fig.1b is a schematic drawing of a GS functionalized noncovalently by TBA and PL-PEG-amine, which form homogeneous and stable suspension in DMF. The GS obtained were shown previously to be highly conducting (~100 fold more conducting than reduced graphene oxide) with much lower degree of oxidation than commonly used graphite oxide [14]. Atomic force microscopy image (Fig. 1a) showed clean sheets of graphene on silicon substrate. Height profile (see Supplementary Material Fig. S1) analysis revealed that GS was ~0.8-1 nm above substrate, corresponding to single layer graphene [12-14], as confirmed by the single symmetrical 2D peak in Raman spectra of individual GS (Fig. 2l and Supplementary Material Fig. S2a).

Surface assembly of our GS on gold occurred when we simply soaked silicon oxide substrates with micro-fabricated gold patterns in a GS suspension, for 5-10 min (see Experimental). Most of the GS were found to selectively adsorb onto gold surfaces, rather than onto the $SiO_2$ substrate (Fig. 2). Scanning Auger elemental/chemical mapping and imaging was carried out to characterize the patterned GS structures. Carbon scanning Auger images (Fig. 2b, e) overlapped with those of gold (Fig. 2c, f), confirming selective self-assembly of GS onto Au patterns.

To verify that we have achieved patterned assembly of single-layer GS, we carried out Raman mapping of assembled GS on gold arrays. G-band Raman mapping image (Fig. 2j) clearly revealed GS assembly structures, consistent with Auger mapping data. A high D/G (disorder to G band) ratio (Fig. 2k) was due to the small sizes of GS (~400 nm average) compared to the laser spot (~1 μm), which probed defects on the edges of GS. Disorder could also be due to in-plane defects, but was difficult to discern from the edges. The 2D band of the

GS was well fitted by a single Lorentzian peak, indicative of the single-layer nature [18, 19]. In contrast, double-layer graphene showed significantly different 2D band with four superimposed peaks (see Supplementary Material Fig. S2b).

By reducing the sizes of gold patterns to ~150nm, we pushed the limit of GS assembly on Au. That is, putting a single GS on a gold island. Fig. 3a shows an SEM image of GS adsorbed onto small gold islands (~150nm size with ~1.5μm pitch). Many of the Au islands have attracted a single GS. Fig. 3b shows a carbon scanning Auger image and an SEM image side by side. The C-Auger image of GS matched well with the SEM image of GS trapped on gold islands, showing Auger chemical imaging can be used to characterize single GS effectively.

Through a series of control experiments (see Supplementary Material Table. S1), we investigated the underlying GS self-assembly mechanism. Both TBA and the amino group of PL-PEG-amine could interact with Au surface. However, our control experiments suggested that the latter was not critical. GS functionalized by TBA and PL-mPEG (with methyl terminal group) showed similar selective adsorption on Au (see Supplementary Material Fig. S3b), while GS with TBA removed by washing rarely adsorbed onto Au surface (see Supplementary Material Fig. S3c). We also tested GS assembly on other metals including Al, Ti, Co and Pd. None of these metals showed as good attracting effects to GS as Au (see Supplementary Material Fig. S5). This is consistent with the strong electrostatic interactions between TBA and negatively-charged gold in the literature [20, 21]. Our GS/DMF suspension contained ~3% water by volume, which facilitated the electrostatic self-assembly mechanism. No adsorption on Au could be observed if we transferred GS into pure DMF (see Supplementary Material Fig. S4b). Moreover, hydroxide anions in aqueous phase were known to adsorb onto Au surface to form a negatively charged layer [22, 23], which then attracted the positively charged GS/TBA for surface assembly. This was further verified by our control experiment in which we acidified the GS suspension with nitric acid (pH decreased from 9.5 to 4) to neutralize the hydroxide ions (nitric ions were not known to adsorb strongly on Au), and the adsorption of GS onto Au surface was significantly reduced (see Supplementary Material Fig. S6).

The selective adsorption of GS on gold opened up a new way to build graphene devices. We soaked substrate with pre-formed Au source-drain (S-D) electrode arrays into a GS suspension (see Experimental), the noncovalent self-assembly process navigated the TBA functionalized GS onto gold electrodes, with some of the GS bridging two closely spaced Au electrode fingers (Fig. 3c). This method afforded an efficient way to fabricate GS electrical devices at a large scale. Typically, we made ~100 device patterns on a $0.5\times0.5cm^2$ chip, and 10-20% of them turned out to be single-connection GS devices.

Fig. 4a shows an AFM image of a typical single GS device fabricated by the self-assembly method. The device was comprised of a $p^{++}$-Si backgate, 100nm $SiO_2$ as gate dielectrics and Ti(2nm)/Au(10nm) as source-drain electrodes (for bottom contacts). All fabricated GS devices were probed in a vacuum probe station with a base pressure of $\sim2\times10^{-6}$ Torr. The as-made device showed almost no gate dependence with high resistance (Fig. 4c, red curve). By sweeping to high source drain bias of ~4V, we observed that current through the GS showed a sudden jump, with a large hysteresis upon sweeping back the bias voltage (Fig. 4b). The conductance of the device increased significantly (Fig. 4c, blue curve) after this high current treatment cycle, which was attributed to electrical annealing and heating effects to the GS and bottom contacts. It was previously reported that high current annealing could heat up peeled-off graphene devices to remove contaminants and make shift Dirac point (DP) of graphene to near zero gate-voltage [24]. Similar effects were observed with our chemically derived GS. After electrical annealing, a clear DP appeared around $V_g$~0 V, indicating very low doping level. Importantly, AFM imaging found that the electrical annealing process removed coating molecules on GS, with a ~0.3-0.5nm decrease in the apparent height or thickness of the GS after electrical annealing (see Supplementary Material Fig. S8). Thus, the electrical-annealing method removed the molecules on the GS responsible for the self-assembly process and the associated doping effect, leading to graphene sheet devices exhibiting clear DP that are intrinsic to pristine graphene.

We also obtained fully-suspended graphene sheet device by self-assembly of GS between tall metal electrodes. We shortened the source-drain electrode distance to less than 100 nm and increasing the electrode height to ~50 nm. Graphene sheets were able to adsorb and bridge the electrodes via self-assembly and became suspended. This simple solution processing method readily leads to suspended GS device (Fig. 4d) exhibiting DP after electrical annealing (Fig. 4f) without using sophisticated fabrication/etching methods [25-27].

Our GS was synthesized by a mild process without extensive covalent functionalization of the GS [14] and electrical annealing was used to remove molecular coating and increase the conductance of the GS devices. The average resistivity (defined as $\frac{R \times W}{L}$, where $R$ is resistance of device, $W$ and $L$ indicating GS width and channel length) at the minimum conductance point (MCP) was ~30 kΩ after electrical annealing (Fig. 4g), about 100 times lower than commonly used graphene oxide (GO) in reduced states [12, 13, 15]. Our GS exhibited ~2-5 times higher resistivity than pristine peel-off graphene [24, 28] (Fig. 4g), suggesting small disorder in the plane of the GS, which could also contributed to the D/G ratio in Raman spectroscopy. Nevertheless, our GS devices are ~100 times more conducting than reduced GO with clear DP similar to pristine graphene. Such devices are useful for various applications, as demonstrated by ammonia detection at ppb level (Fig. 4h) [8, 29, 30]. Ammonia caused a clear shift in the DP of our GS as a result of n-doping, which served as the basis for molecular detection unattainable with graphene devices without Dirac points and little gate dependence of electrical conductance.

**Experimental**

GS making method used was developed from our previous work [14]. Typically, 20 mg natural graphite flakes were ground with NaCl crystallites. After removing NaCl by washing with water and filtration, the obtained solid was soaked in 10 ml oleum (with 20% free $SO_3$) and 1 g $NaNO_3$, stirred for 1 day. The acid and salt were washed repeatedly and thoroughly by water, and then the obtained solid was dispersed in 10ml dimethylformamide (DMF), to which 300 μl of tetrabutylammonium (TBA) hydroxide aqueous solution (40%) was added. After 10 minutes of cup-horn sonication, the suspension was kept still for 2 days, and then 4 ml was taken out and bath-sonicated with 15 mg phospholipid-polyenthyleneglycol-amine (PL-PEG-$NH_2$, M.W. ~5,000) for 1 h, forming a homogeneous suspension. After 15000rpm centrifuge for 3min, a black homogeneous supernatant was obtained.

Au patterns on $SiO_2$-coated (~100 nm) Si substrate were fabricated through a procedure of electron beam lithography (EBL), metallization and lift-off. The patterned substrate was soaked in GS suspension for 5-10 minutes, and then it was rinsed by $H_2O$, 2-propanol, and dried by Argon flow.

The GS devices were obtained in similar way. A typical on-substrate device was comprised of a $p^{++}$-Si backgate, 100nm $SiO_2$ as gate dielectrics and Ti(2nm)/Au(10nm) as source-drain electrodes. Pd(40nm)/Au(10nm) were used as source-drain electrodes for fully suspended GS device fabrication. After electrode fabrication, the chip was soaked in GS suspension for ~5 minutes, and then rinsed by $H_2O$, 2-propanol, and dried by Argon flow. All fabricated GS devices were probed in a vacuum probe station with a base pressure of ~$2\times10^{-6}$ Torr.

AFM images were taken of the GS with a Nanoscope IIIa multimode instrument. Scanning Auger elemental mapping was carried out with PHI 700™ scanning auger nanoprobe. The mapping images were plotted using the ~275eV (KLL) peak for carbon, and the ~75eV (MNN) peak for gold. Raman mapping images and spectra were taken with HORIBA JOBIN YVON Raman spectrometer (633nm laser excitation). SEM images were taken with Raith150 nanofabrication system. Electrical characteristics of fabricated devices were measured with Hewlett Packard 4156B precision semiconductor parameter analyzer.

**Acknowledgements**

This work is supported by MARCO-MSD, ONR and Intel.

**Electronic Supplementary Material:**

Further characterizations of GS and self-assembly structures with AFM, Raman and Auger can be found in the supplementary material with 9 supplementary figures.

**Figures**

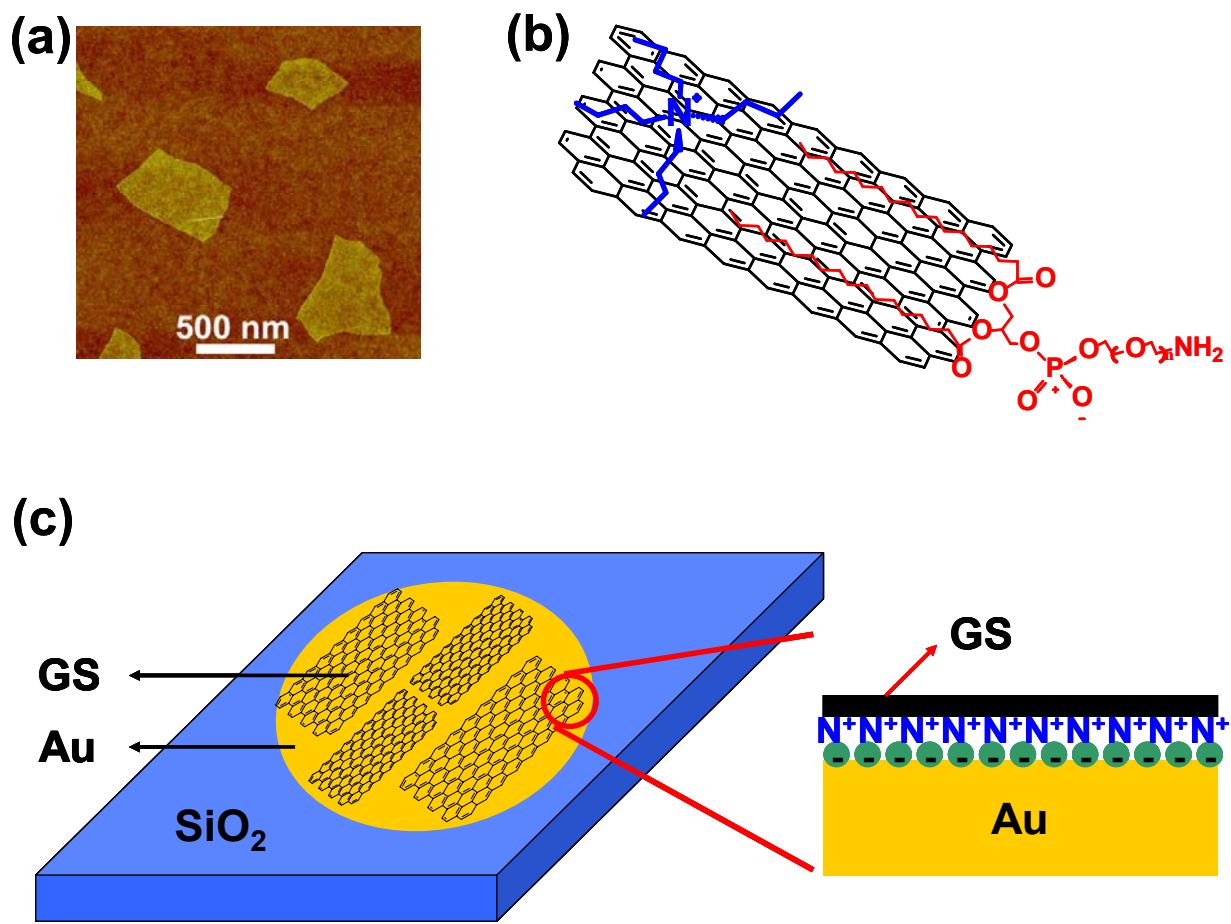

**Figure 1** Self assembly of graphene sheets (GS) on gold. (a) An AFM image of as-synthesized GS. (b) A schematic drawing of non-covalently functionalized GS. (c) A schematic drawing of selective adsorption of GS on a gold pattern on silicon dioxide, mediated by electrostatic interactions between positively charged groups on GS and negative ions adsorbed on Au (right panel).

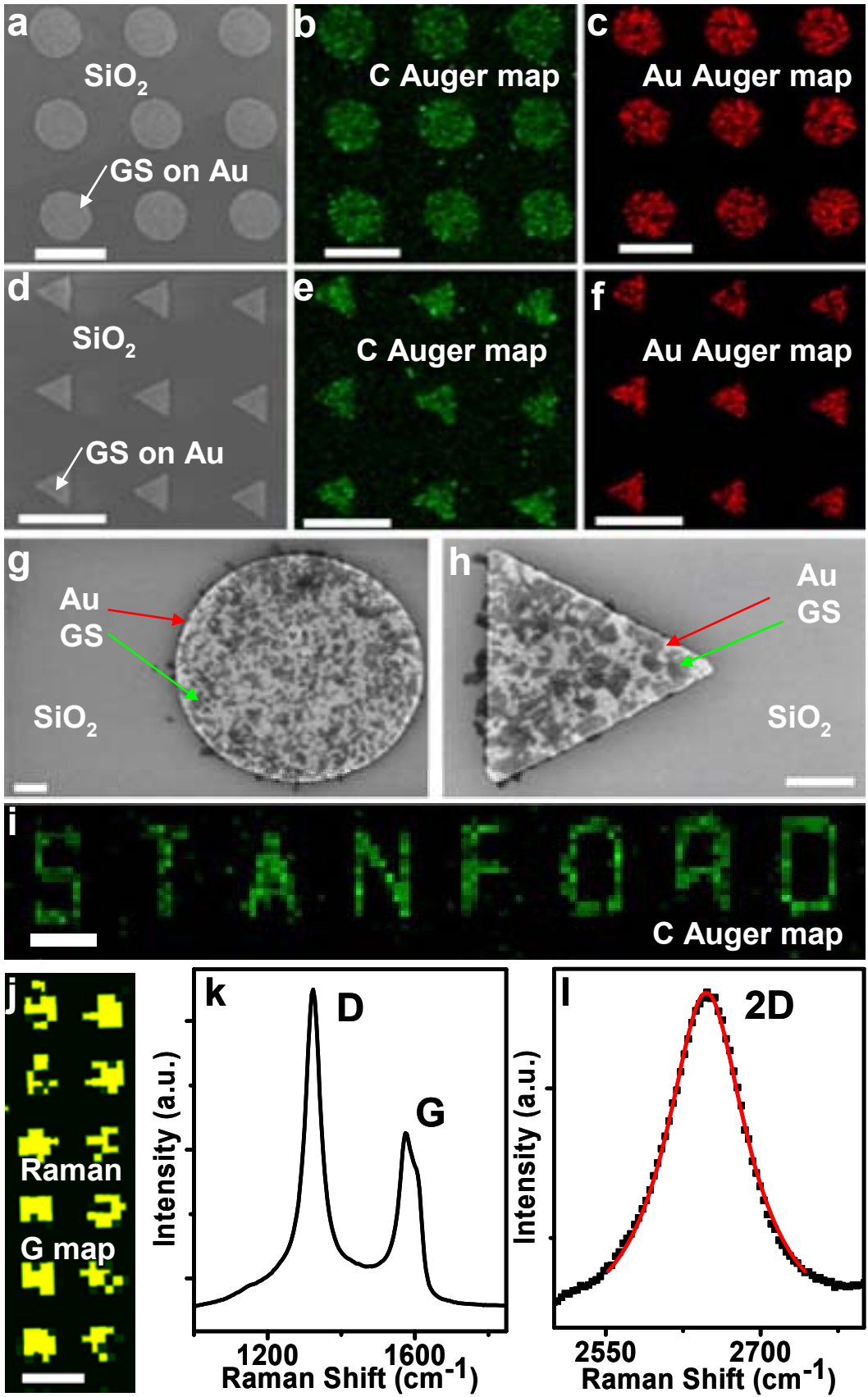

**Figure 2**  Self-assembled GS on gold structures, characterized by scanning electron microscopy, scanning Auger microscopy, and Raman spectroscopy. (a) and (d): SEM images. (b) and (c): Auger images for carbon and gold elements, respectively. (e) and (f): Auger images for carbon and gold elements, respectively. (g) and (h): SEM images showing GS adsorbed onto gold structures. The dark structures in the gold islands were self-adsorbed graphene sheets GS (sheet size ~200nm-1μm). (i) Carbon scanning Auger image of a word 'STANFORD' made of patterned Au lines with GS selectively adsorbed onto the word. That is, graphene spells out 'STANFORD'. (j) Raman mapping (G-band) image of assembled GS arrays on Au. (k) and (l): Raman spectrum of assembled GS on Au structures, in (l), dotted line is Raman 2D band, and red solid line is a single-Lorentzian fit. Scale bars: 10μm (a-f, i, j), 1μm (g, h).

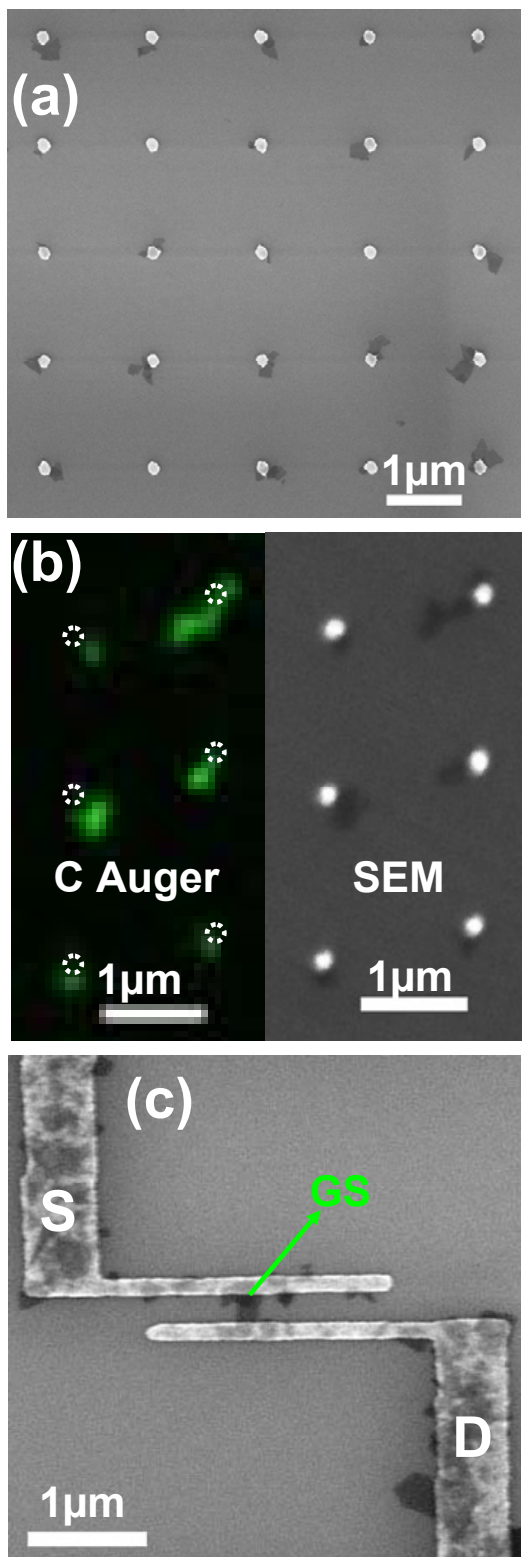

**Figure 3** Single GS arrays and self-assembled GS devices. (a) An SEM image of small gold islands (~150nm) trapping ~1 GS on each island. The dark structures attached to the gold islands were self-adsorbed graphene sheets GS. (b) Side-by-side carbon scanning Auger image (left, white dotted circles were drawn to indicate the gold positions) and SEM image (right, The dark structures were GS.) of GS trapped on gold arrays, (c) A SEM image of GS adsorbed on gold electrodes with a single GS bridging two electrode fingers.

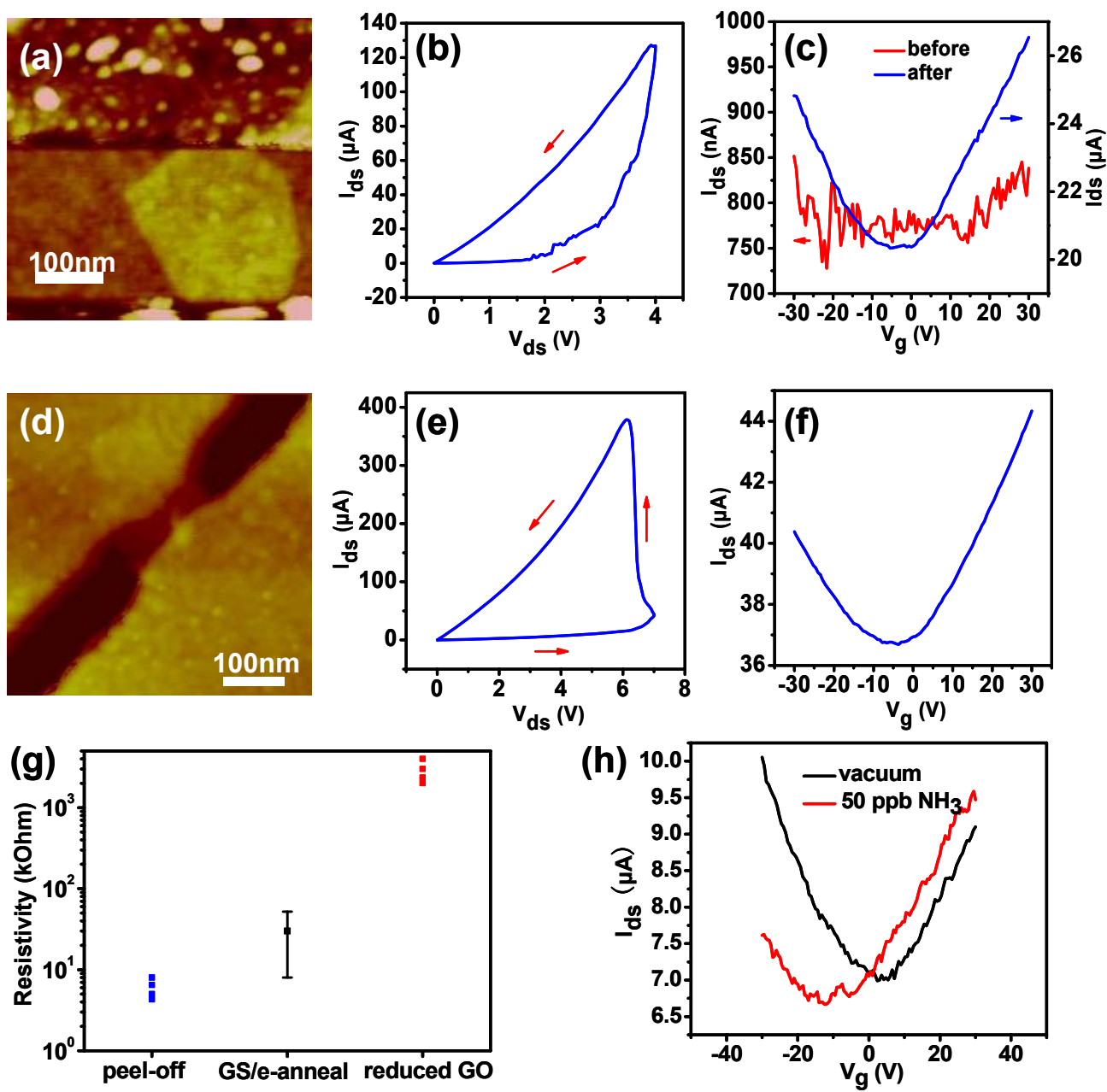

**Figure 4** Electrical devices of self-assembled graphene on-substrate and in fully suspended form. (a) An AFM image of a single GS bridging two Au electrodes on a substrate. (b) Electrical-annealing current-voltage curves recorded by sweeping the bias voltage across the GS. The red arrows show the direction of the voltage sweeping. (c) Current-gate voltage curves ($V_{ds}$=1V) of the same GS device before and after electrical-annealing. (d) An AFM image of a suspended GS device. (e) Electrical-annealing current-voltage curves recorded by sweeping the bias voltage across the GS in (d). (f) Current-gate voltage curves ($V_{ds}$=1V) of the same GS device before and after electrical-annealing. (g) Resistivity comparison of different graphene devices. The resistivity value of electrically annealed GS devices was average of more than 30 devices, with the error bar indicating standard deviation. The multiple resistivity values of peel-off graphene and reduced GO were based on electrical transport data in the literature (see Supplementary Material). (h) Current-gate voltage curves ($V_{ds}$=1V) of a single GS device before and after exposure to 50 ppb ammonia gas.